\documentclass[
    ,final            % use final for the camera ready runs
  ]
  {aipproc}

\layoutstyle{8x11double}

\begin{document}

\title{{Neutrino Scattering on Glass:  NuSOnG} }

\classification{12.15.-y,12.15.Ji,12.15.Mm,12.38.Qk,12.60.Cn,13.15.+g}

\keywords{neutrino, electroweak, beyond standard model, parton distribution functions}

\author{J.M. Conrad, for the authors of the NuSOnG EOI}{
  address={Columbia University, New York, NY 10027}
}

\begin{abstract}

  These proceedings describe the physics goals and initial design for
  a new experiment: NuSOnG -- Neutrino Scattering On Glass.  The
  design will yield about two orders of magnitude higher statistics
  than previous high energy neutrino experiments, observed in a
  detector optimized for low hadronic energy and electromagnetic
  events.  As a result, the purely weak processes $\nu_{\mu}+e^-
  \rightarrow \nu_{\mu}+ e^-$ and $\nu_{\mu}+ e^- \rightarrow \nu_e +
  \mu^-$ (inverse muon decay) can be measured with high accuracy for
  the first time.  This allows important precision electroweak tests
  and well as direct searches for new physics. The high statistics
  also will yield the world's largest sample of Deep Inelastic (DIS)
  events for precision parton distribution studies.

\end{abstract}

\maketitle

%%%%%%%%%%%%%%%%%%%%%%%%%%%%%%%%%%%%%%%%%%%%
%% MAINMATTER
%%%%%%%%%%%%%%%%%%%%%%%%%%%%%%%%%%%%%%%%%%%%

This talk summarized the motivation for a new high energy, ultra-high
statistics neutrino experiment at Fermilab: NuSOnG (Neutrino
Scattering On Glass).  The idea for this experiment arises from the
work of 27 physicists \cite{EOIauthors} who are the authors of an
Expression of Interest \cite{EOI} (EOI), submitted to the Fermilab
directorate.  The high statistics and high energy of NuSOnG leads to
wide-ranging physics opportunities that fall into three broad
categories: 1) Indirect searches for new physics at the Terascale, 2)
Direct searches for new physics at GeV energies and 3) Studies of
parton distributions and nuclear effects.  Here we provide some
examples of what can be accomplished in each area.

\begin{figure}[t]
\centering
\scalebox{0.7}{\includegraphics{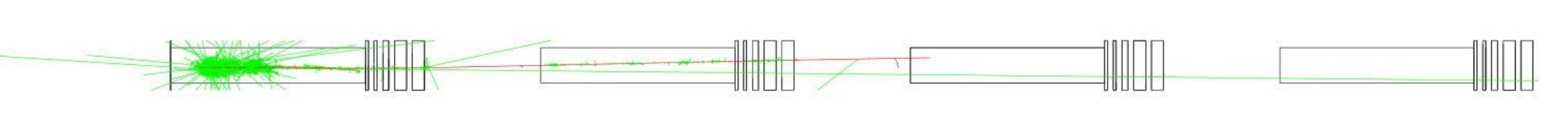}}
\caption{An event display showing a typical deep inelastic event in the 
NuSOnG detector.  The detector consists of four subdetectors, each 
consisting of a glass-based sampling calorimeter followed by an iron-toroid
muon spectrometer.  The event shown here was generated by a 336 GeV 
incoming neutrino, and produced a 218 GeV outgoing muon in a CC DIS 
interaction.}
\label{fig:detector}
\end{figure}

\begin{figure}[t]
{\includegraphics[width=3.in, bb=0 400 560 670]{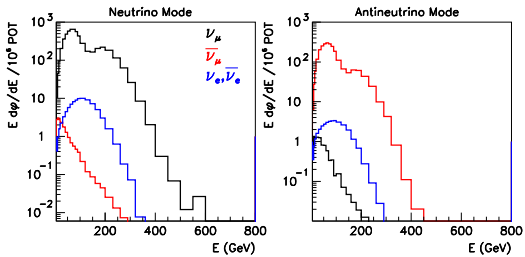}}
\caption{Initial assumption for the NuSOnG flux in neutrino mode (left)
  and antineutrino mode (right).  Black: muon neutrino flux, red: muon
  antineutrino flux, blue: electron neutrino and antineutrino flux. 
  This flux was produced using the NuTeV beam Monte Carlo.}
\label{fig:beam}
\end{figure}

The beam and detector design marry the best aspects of the NuTeV
\cite{NuTeVbeam} and Charm II \cite{CharmIIdet} experiments.  We
propose a 3500 ton (3000 ton fiducial volume) $\mathrm{SiO}_2$
neutrino detector with sampling calorimetry, charged particle
tracking, and muon spectrometers.  The detector consists of four
identical subdetectors, which are separated by regions allowing exotic
particle decay and also calibration beams.  The design is illustrated
in Fig.~\ref{fig:detector}, which shows a GEANT4 \cite{G4} simulation
of a deep inelastic (DIS) interaction.  This detector would run as a
part of a Tevatron Fixed Target Program in the mid-2010's.  The initial
neutrino energy distribution is identical to the NuTeV experiment
\cite{NuTeVbeam}, as shown in Fig.~\ref{fig:beam}.  A challenging
technical aspect of the experiment is the the required Tevatron
protons-on-target (POT) rate, which is 4 $\times$ 10$^{19}$ POT/year
\cite{SyphersMemo}.

An important
aspect of achieving these goals is that the neutrino flux can be well
measured.  The energy dependence of the flux is well determined by the
``low y'' method used by many neutrino experiments \cite{NuTeVflux,
MinosFlux}.  Because the flux lies above 12 GeV, which is the
threshold for inverse muon decay (IMD), the IMD events, which are
well-predicted by the Standard Model, can provide the normalization.
This is the first experiment to have sufficient IMD events to be able
to use this method to obtain a precision measurement of the absolute
flux.  We assume a flux error of 0.5\%.

As with any experiment in the design phase, the NuSOnG group is
exploring options for run-modes.  At NuFACT07 and in the EOI, we
presented a plan which obtained equal statistics (20k events each) for
$\nu$-electron and $\bar \nu$-electron scattering.  In the interim, we
have found that a stronger physics case, for the same POT, is made
with substantially more running in neutrino than antineutrino mode.
In this proceedings, expectations are reported assuming $1.5\times
10^{20}$ protons on target in neutrino mode and $0.5\times 10^{20}$
protons on target in antineutrino mode.  This yields:
\begin{table}[h]
\begin{tabular}{c|c}
600\/M & $\nu_\mu$ CC Deep Inelastic Scattering\\
190\/M & $\nu_\mu$ NC Deep Inelastic Scattering \\
75\/k & $\nu_\mu$ electron NC elastic scatters \\
700\/k &$\nu_\mu$ electron CC quasielastic scatters (IMD) \\
33\/M &  $\bar \nu_\mu$ CC Deep Inelastic Scattering \\
12\/M & $\bar \nu_\mu$ NC Deep Inelastic Scattering \\
7\/k & $\bar \nu_\mu$ electron NC elastic~scatters \\
0\/k &  $\bar \nu_\mu$ electron CC quasielastic scatters \\
\end{tabular}
\end{table}

\section{Indirect Searches for New Physics at TeV Scales}

Indirect searches are those where new physics is identified by
comparing NuSOnG measurements to those from other experiments. As one
example, the
EOI describes precision measurement of the NC couplings, which, when
compared with the LEP invisible $Z$-width measurements, can open a
unique window on new physics.  Here we consider the ``classic
example'' for neutrino scattering --- comparison of electroweak
measurements of $\sin^2 \theta_W$.  

At present, a 3$\sigma$ deviation is observed between measurement of
$\sin^2 \theta_W$ in neutrino DIS scattering from NuTeV \cite{NuTeVEW}
and the LEP/SLD e$^+$e$^-$ results \cite{EWWG}.  This discrepancy is consistent
with past measurements from neutrino experiments, which show a
systematic shift from the Standard Model. However the small errors of
NuTeV make this result much more significant.  The NuTeV electroweak
measurement was presented at NuFACT07 by Kevin McFarland, and will be
described in detail in his contribution to these proceedings.

NuSOnG will measure $\sin^2\theta_W$ in two ways: through the ratio of
$\nu_\mu$-electron elastic scattering ($\nu$eES) to IMD events and
through the Paschos-Wolfenstein (PW) technique which exploits ratios of DIS
NC and CC scattering.   The former technique, which has the virtue of
being a purely leptonic measurement, is new.  Past neutrino-electron
scattering experiments measuring $\sin^2 \theta_W$ have been at low
energies and thus could not normalize to IMD.  The latter
technique was employed by NuTeV.  NuSOnG expects to improve on the
experimental errors which were published by NuTeV by about a factor of
two \cite{EOI}, with much of this improvement coming from the increased
statistics. 

 NuSOnG will also be able to address the two most viable
``standard model'' explanations of the NuTeV anomaly: the strange
sea asymmetry and isospin violation \cite{McFarland}.  The strange sea
asymmetry will be constrained by accurate measurement of dimuon
production in $\nu$ and $\bar \nu$ running modes, as well as an {\it
  in-situ} emulsion-based measurement of the semi-leptonic branching
ratio to charm.  The level of isospin violation can be constrained by
the high-statistics measurement of $\Delta xF_3$ (discussed below).

The NuSOnG experiment provides complementary information to LHC.  Rather than
generalize, to illustrate the power of
NuSOnG, two specific examples are given here.  We emphasize that these
are just two of a wide range of examples, but they serve well to
demonstrate the point.  

\begin{figure}[t]
\centering
{\includegraphics[width=2.25in, bb=100 100 650 500]{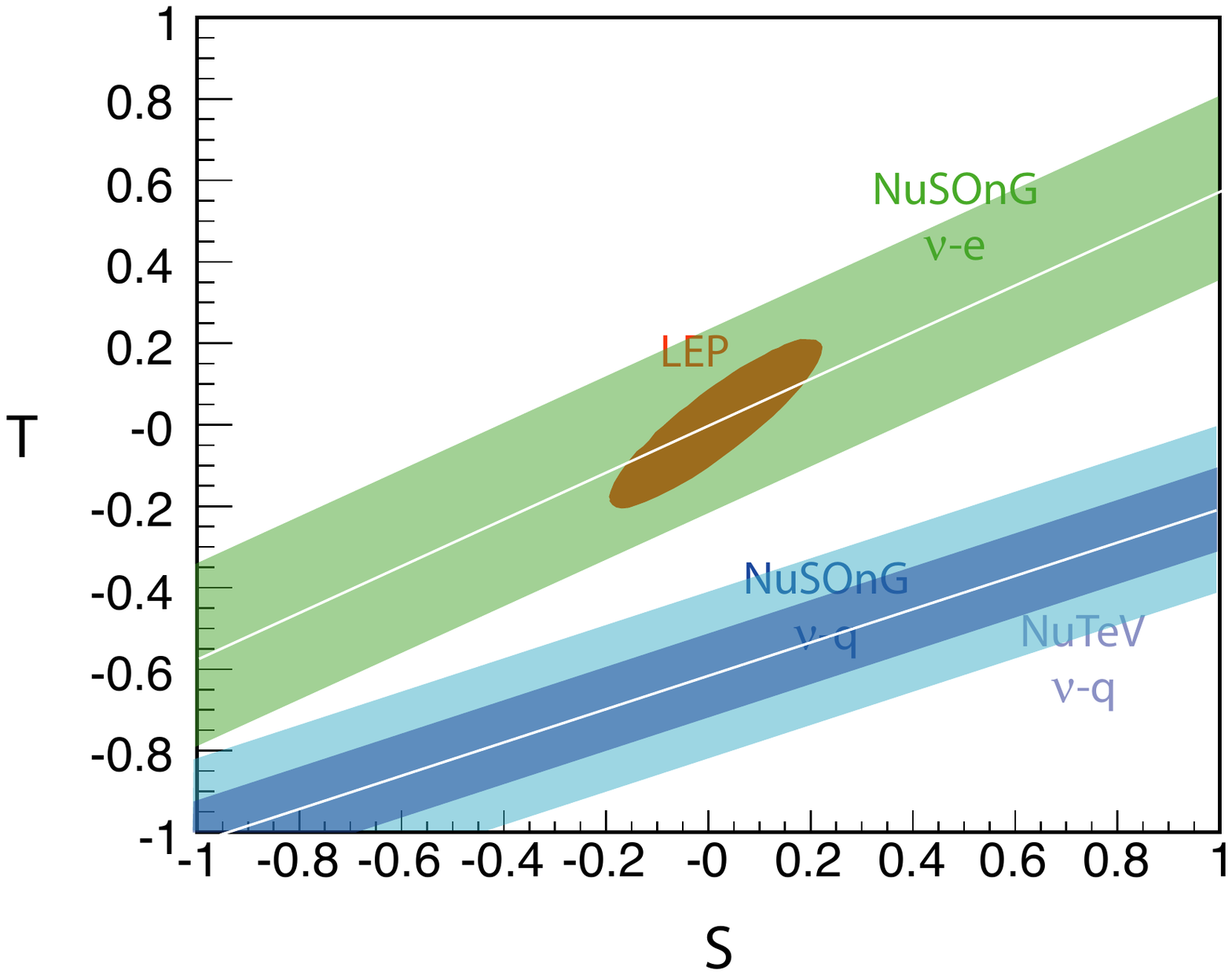}}~~~~~~~~~~~
{\includegraphics[width=2.25in, bb=100 107 650 500]{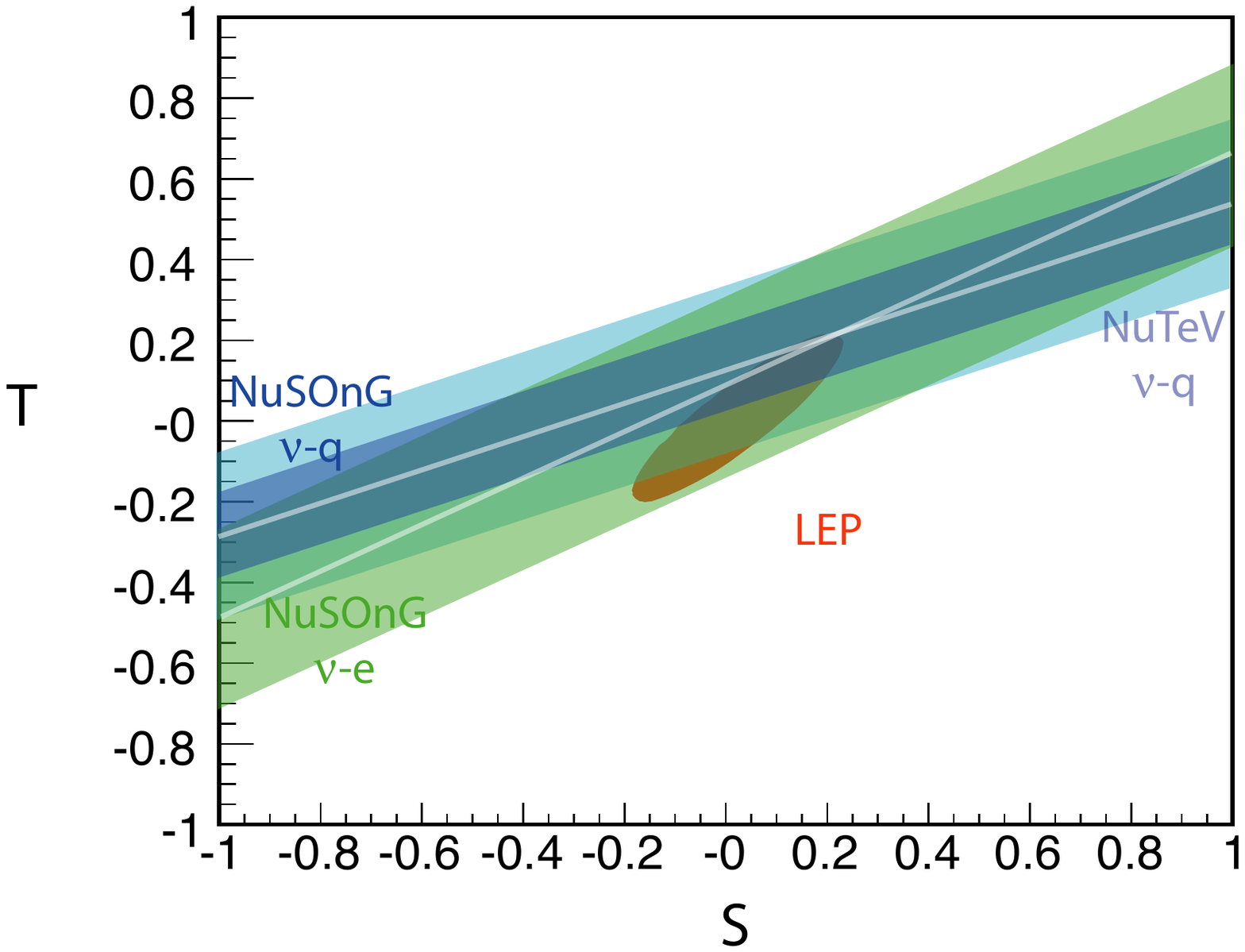}}
\caption{Left: NuSOnG expectation for a 3 TeV $Z^\prime$ which couples
  via $B-3L_\mu$; Right: NuSOnG expectation if the NuTeV anomaly is
  due to either isospin violation or a strange sea asymmetry, and
  there is a heavy 4th generation with isospin violation.}
\label{fig:NuSOnGEW}
\end{figure}

First, consider a heavy $Z^\prime$ which is in the $B-xL$ family -- a
case of interest because this is an anomaly free extension of the
Standard Model \cite{CarenaTait}.  We will consider a 3 TeV $Z^\prime$
which couples to $B-3L_\mu$.  This is one explanation of the
NuTeV anomaly \cite{TakLoinaz, Davidson}.  In this case, the LHC will
see  $Z^\prime \rightarrow \mu \mu$ channel, and measure
and $A_{FB}$, but not the width, due to resolution. The absence of the
$ee$ channel will be clear but absence of the $\tau\tau$ channel will
only be surmised after very high statistics are obtained. Among the
quark channels, the one which is reconstructable is $tt$.  In this
scenario, NuSOnG would find that isopin and the strange sea can be
constrained to the point that they do not provide an explanation for
the NuTeV anomaly, thus NuTeV is the result of new physics. The NuSOnG
PW measurement of $\sin^2 \theta_W$ will agree with NuTeV, and the
$\nu$eES measurement will agree with LEP.  Fig.~\ref{fig:NuSOnGEW}(left)
illustrates this example.  The complementary information from NuSOnG
is needed to narrow the options to the $B-3L_\mu$ coupling.

A second example is the existence of a fourth generation family.  A
fourth family with non-degenerate masses ({\it i.e.} isospin
violating) are allowed within the LEP/SLD constraints \cite{Tait}.  As
a model, we choose a fourth family with mass splitting on the order of
$\sim 75$ GeV and a 300 GeV Higgs. This is consistent with LEP at
1$\sigma$ and perfectly consistent with $M_w$, describing the point
(0.2,0.19) on the ST plot\cite{kribs}.  In this scenario, LHC will
measure the Higgs mass from the highly enhanced $H \rightarrow ZZ$
decay \cite{Tait}. An array of exotic decays which will be difficult
to fully reconstruct, such as production of 6 W's and 2 b's, will be
observed at low rates.  The expected NuSOnG result in this scenario is that the
strange sea or isospin violation will explain the NuTeV anomaly, and
that the corrected NuTeV PW result will agree with the NuSOnG PW and
$\nu$eES measurements. These three precision neutrino results, all
with ``LEP-size'' errors, can be combined and will intersect the
one-sigma edge of the LEP measurements.
Fig.~\ref{fig:NuSOnGEW}(right) illustrates this example.  From this, the
source, a fourth generation with isospin violation, can be
demonstrated.

\section{Direct Searches for New Physics}

NuSOnG is also designed to perform a range of direct searches for new
physics.  The segmented detector design, which was driven by the need
to bring a calibration beam into each segment of the detector, also
allows for a decay region in which one can search for decays of
neutral heavy leptons in the mass ranges of a 10's of MeV to
multi-GeV.  The design of the detector also allows for direct searches
for new physics through neutrino interactions in the detector which
are outside of the SM prediction.  Two examples of this, which are
considered here in detail, are wrong-sign inverse muon decay and
observation of an excess of very high energy $\nu_e$ interactions.

In considering the list of processes which NuSOnG can observe (see
above), ``wrong-sign IMD'' (WSIMD) was explicitly listed as resulting
in zero events.  In the Standard Model, the interaction $\bar \nu_\mu
+ e^-$ cannot occur, since it violates lepton family number
conservation ($\Delta L_e = -\Delta L_\mu = 2$).  
A number of theories beyond the Standard Model predict that lepton
flavor number is not a true conserved quantum number; this means that
processes that violate lepton flavor are allowed to occur.  Theories
which incorporate multiplicative lepton number
conservation~\cite{Feinberg, Ibarra:2004pe}, left-right
symmetry~\cite{Herczeg:1992pt}, or the existence of
bileptons~\cite{Godfrey:2001xb} fall under this category.

In principle, the observation of a single muon with no hadronic energy
in antineutrino mode is a direct signature of new physics.  In
practice there are backgrounds from $\nu_\mu$ contamination in the
$\bar \nu_\mu$ beam, $\nu_e$ contamination and charge
misidentification of candidate muons.  With these backgrounds in mind,
NuSOnG is designed to improve the limit on WSIMD by an order of
magnitude from the present level\cite{Formaggio:2001jz}.  If we assume a
conservative knowledge of the backgrounds at the 5\% level, this would
imply a limit on the lepton number violation cross-section ratio of
better than 0.2\% (at 90\% C.L.) for V-A couplings and less than
0.06\% for scalar couplings.  Previous searches, based on $1.6 \times
10^{18}$ protons on target and smaller target masses, have placed
limits on this cross-section ratio to less than 1.7\% at 90\% C.L. for
V-A couplings and less than 0.6\% for scalar
couplings~\cite{Formaggio:2001jz}.

In fact, the best argument for searching for new lepton flavor
violating effects arises from the experimental observation of neutrino
oscillations, which explicitly violates lepton flavor.  This describes
the conversion of neutrino flavors as a function of time via a $3
\times 3$ mixing matrix.  In most analyses, this matrix is assumed to
be unitary.  However new physics at high energy scaales 
can induce nonunitarity in this
matrix.  

Nonunitarity, or ``Matrix Freedom,''
introduces striking changes to the probability formula for
neutrino flavor transitions.   
The level at which unitarity is violated can be defined as $X_{\alpha}$, where 
\begin{equation}
\label{eq:unitarityviol}
\sum_j |U_{\alpha j}|^2 = 1-X_\alpha,
\end{equation}
with $X_{\alpha}$ being small.  One of the main consequences of such
a scenario is instantaneous ($L=$0) flavor transitions in a neutrino
beam.  Extending the argument of ref.~\cite{hep-ph/0607020}, the
non-orthogonality of $\nu_{\mu}$ and $\nu_e$ results in an
instantaneous transition at $L=0$ from $\nu_\mu$ to $\nu_e$
\cite{BorisPrivateComm}.  Thus one could observe an excess of $\nu_e$
events in a pure $\nu_{\mu}$ beam.  Similarly, 
a $\bar \nu_\mu$ to $\bar \nu_e$ transition at $L=0$, then a
subsequent $\bar \nu_e + e^- \rightarrow \bar \nu_\mu \mu^-$ interaction will
produce a WSIMD signal in NuSOnG.

From
\cite{hep-ph/0607020}, the limits on
$\nu_\mu \rightarrow \nu_e$ instantaneous transition are at the $\sim
1\times 10^{-4}$ level, and arise from physics above the EW scale
which would be apparent in corrections to decay rates.  This is
at the edge of NuSOnG capability, but may be observable depending on
the control of the systematics.   However, these limits are not applicable
if the nonunitarity arises due to an effect such as the existence of
a ``neutrissimo'' -- a $\sim 100$ GeV neutral heavy lepton -- which 
mixes with the light neutrinos and thereby affects the apparent coupling.
In this case, NuSOnG has substantial allowed range for its search.
Note that this WSIMD signature can only be
observed in a high energy beam such as at NuSOnG because the
threshold for muon production is 12 GeV.

An alternative method to search for this effect is to look for $\nu_e$
appearance in an energy range with low, and well-constrained,
intrinsic $\nu_e$ background.  In the case of NuSOnG, this is on the
high energy tail of the flux, above $\sim 200$ GeV.  For the $\sim
1\times 10^{-4}$ level limit on $\nu_\mu$ transformation to $\nu_e$
quoted in \cite{hep-ph/0607020}, NuSOnG would see an excess of
$\sim200$ $\nu_e$ events in this high energy region and a 10\%
increase in flux for E$\sim350$ GeV.  In that region, the
$\nu_e$ flux is mainly from $K^+$ decay, which is well constrained by
the $K^+$-produced $\nu_\mu$ events. Such an excess should therefore be
straightforward to observe.  An observation of this excess in both
this mode and as WSIMD would be a very striking signature for this
effect.

\section{Parton Distribution Studies}

NuSOnG also will study parton distributions and
nuclear effects to high precision.  Comparison of our resut to 
the charged lepton scattering data can provide clues to the sources of the
major features which appear in nuclear effects: shadowing,
antishadowing, and the EMC effect.  Present data 
suggest nuclear corrections for the $\nu \ \mbox{and}\
\overline\nu$ cross sections which differ from expectation~\cite{CTEQ, kp}.
In order to investigate this, NuSOnG will include targets
of C, Al, Fe, and Pb, as well as SiO$2$. This study will
complement results of Miner$\nu$a and of eRHIC.

The high statistics and isoscalarity allows measurement of structure
function combinations which are so-far poorly constrained. Among these
is the precision measurement of $\Delta xF_3
=xF_{3}^{\nu}-xF_{3}^{\bar{\nu}}$, which provides sensitivity to
isospin violation.  The sensitivity arises from residual
$u,d$-contributions.  The effect is amplified compared to the $s$ and
$c$ contributions because $d\rightarrow u$ transitions are not subject
to slow-rescaling corrections which strongly suppress the
$s\rightarrow c$ contribution to $\Delta
xF_{3}$. \cite{Kretzer:2001mb}.  The ability of NuSOnG to separately
measure $xF_{3}^{\nu}$ and $xF_{3}^{\bar{\nu}}$ over a broad kinematic
range, and with better high-$y$ acceptance than NuTeV/CCFR, will provide
powerful constraints on the sensitive structure function combination
$\Delta xF_{3}$.

\section{Summary}

These proceedings have covered some highlights of the measurements
which could be made at a new high-statistics, high-energy neutrino
scattering experiment, called NuSOnG.  The goal is to obtain a high 
statistics sample of well reconstructed $\nu_\mu$ electron scatters,
as well as $\nu$DIS events, leading to physics opportunities in 
the area of electroweak precision measurement, direct searches
and QCD studies.

\end{document}